\documentclass[proof]{pasj01}
\Received{$\langle$reception date$\rangle$}
\Accepted{$\langle$acception date$\rangle$}
\Published{$\langle$publication date$\rangle$}
\usepackage{microtype}

\begin{document}

\title{Magnetic Field Analysis of the Bow and Terminal Shock of the SS433 Jet}
\author{Haruka \textsc{Sakemi}$^{1,*}$, \and Mami \textsc{Machida}$^{1}$, \and Takuya \textsc{Akahori}$^{2,3}$, \\ \and Hiroyuki \textsc{Nakanishi}$^{2}$, \and Hiroki \textsc{Akamatsu}$^{4}$, \and Kohei \textsc{Kurahara}$^{2}$, \and \\and Jamie \textsc{Farnes}$^{5,6}$}%
\altaffiltext{1}{Graduated School of Science, Kyushu University, 744 Motooka Nishi-ku, Fukuoka, Fukuoka 819-0395, Japan}
\altaffiltext{2}{Graduated School of Science and Engineering, Kagoshima University, 1-21-35 Korimoto, Kagoshima, Kagoshima 890-0065, Japan}
\altaffiltext{3}{National Astronomical Observatory of Japan, 2-21-1 Osawa, Mitaka, Tokyo 181-8588, Japan}
\altaffiltext{4}{SRON Netherlands Institute for Space Research, Sorbonnelaan 2, 3584 CA Utrecht,  the Netherlands}
\altaffiltext{5}{Department of Astrophysics/IMAPP, Radboud University, PO Box 9010, NL-6500 GL Nijmegen, the Netherlands}
\altaffiltext{6}{Oxford e-Research Centre (OeRC), Keble Road, Oxford OX1 3QG, England}
\email{sakemi@phys.kyushu-u.ac.jp}


\maketitle

\begin{abstract}
We report a polarization analysis of the eastern region of W50, observed with the Australia Telescope Compact Array (ATCA) at 1.4 -- 3.0~GHz. In order to study the physical structures in the region where the SS433 jet and W50 interact, we obtain an intrinsic magnetic field vector map of that region. We find that the orientation of the intrinsic magnetic field vectors are aligned along the total intensity structures, and that there are characteristic, separate structures related to the jet, the bow shock, and the terminal shock. The Faraday rotation measures (RMs), and the results of Faraday Tomography suggest that a high intensity, filamentary structure in the north-south direction of the eastern-edge region can be separated into at least two parts to the north and south. The results of Faraday Tomography also show that there are multiple components along the line of sight and/or within the beam area. In addition, we also analyze the X-ray ring-like structure observed with \textit{XMM-Newton}. While the possibility still remains that this X-ray ring is ``real", it seems that the structure is not ring-like at radio wavelengths. Finally, we suggest that the structure is a part of the helical structure that coils the eastern ear of W50.
\end{abstract}

\section{Introduction}
W50 (J2000 RA: $19^{h}$$08^{m}$$10^{s}$  -- $19^{h}$$16^{m}$$50^{s}$, DEC: $04^{\circ}$$30^{\prime}$$00^{\prime\prime}$ -- $05^{\circ}$$30^{\prime}$$00^{\prime\prime}$) is a large radio nebula located near the Galactic plane. W50 exhibits an elongated shell structure in the east--west direction and the protruding regions are called ``ears'' (\cite{Geldzahler1980}; \cite{Downes1981}; \cite{Downes1986}; \cite{Elston1987}; \cite{Dubner1998}). \citet{Dubner1998} produced a high-quality radio image from an observation with the Very Large Array (VLA), and revealed that the eastern ear exhibits a helical pattern in the total intensity image. There is also a filamentary radio structure in the north--south direction of the eastern ear (RA: $19^{h}16^{m}00^{s}$, DEC: $04^{\circ}39^{\prime}00^{\prime\prime}$ -- $05^{\circ}00^{\prime}00^{\prime\prime}$) as seen by observations with the VLA (\cite {Elston1987}; \cite{Dubner1998}). Around the center of W50's main spherical structure, there is a micro-quasar SS433 ejecting a spiral jet with a speed of 0.26 times the speed of light (\cite{Abell1979}; \cite{Eikenberry2001}). The major axis of W50 is well aligned with the SS433 jet, implying that the ram pressure from the jets extended W50 and made the ears as well as the helical pattern and the filament in the eastern ear (\cite{Elston1987}; \cite{Dubner1998};  \cite{Goodall2011a}; \cite{Goodall2011b}).  
\par W50 and SS433 have been observed frequently at not only radio but also optical, X-ray and other wavelengths, and correlations among them have been reported. For example, ROSAT observations found a correlation between the radio and X-ray in the north--south filament of the eastern ear, and suggested that the filament corresponds to the terminal shock of the SS433 jet (\cite{Brinkmann1996}; \cite{Safi-Harb1997}; \cite{Safi-Harb1999}). On the other hand, with \textit{XMM-Newton}, \citet{Brinkmann2007} found an X-ray ring-like structure in the eastern ear, and suggested that this ring corresponds to the terminal shock. With the images of [O$_{\rm I\hspace{-.1em}I\hspace{-.1em}I}$] 5007 \AA\  and H$\alpha$ + [N$_{\rm I\hspace{-.1em} I }$] 6548, 6584 \AA, \citet{Boumis2007} detected another filamentary structure at the intermediate region between the eastern ear and W50's main spherical structure. Near this region, lenticular-shaped X-ray structures are found \citep{Brinkmann2007}, and it is claimed that this structure was made by the interaction between the re-collimated jet and the surrounding matter. There is also faint optical emission at the northern-edge of W50's main spherical structure \citep{Boumis2007}.
\par Meanwhile, polarization observations and analyses have rarely been carried out for W50 and the eastern ear (\cite{Downes1981}; \cite{Downes1986}; \cite{Gao2011}). The detailed magnetic field structure of the ears has not yet been clearly determined. \citet{Downes1981}, (\yearcite{Downes1986}) observed W50 at 1.7, 2.7, and 5 GHz with the Effelsberg 100-m telescope, and reported high polarized fractions at the northern-edge of the spherical structure with magnetic fields oriented parallel to the sphere. They also found an irregular distribution of Faraday rotation measure (RM) in W50, albeit only using two frequencies. Recently, \citet{Farnes2017} performed full polarization, mosaicked observations of the whole region of W50 by using the Australia Telescope Compact Array (ATCA). They carried out multi-wavelength (1.4 -- 3.0 GHz) observations and revealed the distribution of RM. They suggested the existence of a loop magnetic field surrounding the sphere, and that the symmetry axis is perpendicular to the jet axis. It was also suggested that a ring-like magnetic field was present at the north--south filament in the eastern ear. They compared the ATCA images with the H$\alpha$ images obtained from the Isaac Newton Telescope Photometric H$\alpha$ Survey of the Northern Galactic Plane (IPHAS) (\cite{Drew2005}; \cite{Barentsen2014}), and claimed that the depolarization is induced by foreground, ionized, H$\alpha$ emitting gas. They identified new faint optical H$\alpha$ filaments and corresponding radio structures. 
\par Understanding the detailed structure of the ears is essential to know how the jet propagates and how W50 interacts with the jet. It is especially important to know the magnetic field structure, as this is a clue to reveal the connections between W50, the jet, and the surrounding environment. In this paper, using the data from \citet{Farnes2017} we further analyze polarized radio emission from the eastern ear in order to obtain the intrinsic magnetic field vector map. This paper is organized as follows: in Section 2, we briefly review the observations and data reduction. In Section 3, the results are provided. In Section 4, we present our discussion about the eastern ear. In Section 5, we summarize the conclusions.

\section{Observations and Data Reduction}
Throughout this paper, we have used the radio data introduced by \citet{Farnes2017}. We here briefly review the observations and the data reduction. See \citet{Farnes2017} for further details.

\begin{table}[t]
\begin{center}
\caption{Details of the ATCA observations.}\label{tb:1}
\begin{tabular}{ll}
\hline
Parameter & Value \\
\hline
Field of View & $3^{\circ} \times 2^{\circ}$ \\
No. Channels & 2048 \\
Central Frequency & 2.1 GHz \\
Bandwidth & 2.0 GHz \\
Resolution of  image & $3.9 \times 2.9$ arcmin$^{2}$ \\
\hline
\multicolumn{2}{@{ }@{ } } {\hbox to 0pt{\parbox{85mm}\hss}}
\end{tabular}
\end{center}
\end{table}

\begin{table*}[t]
\begin{center}
\caption{Means of total intensity (Stokes $I$), Stokes $Q$, $U$ and polarized intensity in the region RA (J2000): $19^{h}15^{m}02.597^{s}$ -- $19^{h}16^{m}52.863^{s}$, Dec (J2000): $04^{\circ}36^{\prime}26.61^{\prime\prime}$ -- $05^{\circ}03^{\prime}58.80^{\prime\prime}$, the noise values of total intensity (Stokes $I$), Stokes $Q$ and $U$ in the no emission region (RA (J2000): $19^{h}16^{m}49.408^{s}$ -- $19^{h}18^{m}41.506^{s}$, Dec (J2000): $05^{\circ}05^{\prime}08.36^{\prime\prime}$ -- $05^{\circ}56^{\prime}41.87^{\prime\prime}$), and the noise values of polarized intensity estimated by equation (1) at each frequency.}\label{table2}
\begin{tabular}{llllll}
\hline
\multicolumn{3}{c}{Frequency (GHz)} & \multicolumn{1}{c}{3.0} & \multicolumn {1}{c}{2.1} & \multicolumn {1}{c}{1.4}\\ \hline
Total Intensity (Stokes $I$) & Mean (mJy/beam) & 10 MHz channel & 14.70 & 17.92 & 53.81 \\
 & & 100 MHz channel & 12.84 &17.41 & 49.84 \\
 & Noise (mJy/beam) & 10 MHz channel & 2.93 & 12.3 & 12.2 \\
 & & 100 MHz channel & 2.35 & 6.02 & 11.3 \\
 Stokes $Q$ & Mean (mJy/beam) & 10 MHz channel & 0.9110 & 0.5095 & 0.6452 \\
 & & 100 MHz channel & 0.8049 &0.6296 & 0.8638 \\
 & Noise (mJy/beam) & 10 MHz channel & 3.44 & 9.57 & 7.66 \\
 & & 100 MHz channel & 2.08 & 6.07 & 6.55 \\
Stokes $U$ & Mean (mJy/beam) & 10 MHz channel & -0.02145 & 0.5468 & -0.07544 \\
 & & 100 MHz channel & 0.05078 &0.3525 & -0.8491 \\
 & Noise (mJy/beam) & 10 MHz channel & 3.81 & 9.05 & 6.92 \\
 & & 100 MHz channel & 2.78 & 4.48 & 5.95 \\
Polarized Intensity & Mean (mJy/beam) & 10 MHz channel & 8.020 & 11.94 & 10.22 \\
& & 100 MHz channel & 7.072 & 11.26 & 9.433 \\ 
 & Noise (mJy/beam) & 10 MHz channel & 1.53 & 1.53 & 1.53 \\
 & & 100 MHz channel & 0.500 & 0.500 & 0.500 \\
\hline
\multicolumn{6}{@{ }@{ } } {\hbox to 0pt{\parbox{85mm}\hss}}
\end{tabular}
\end{center}
\end{table*}

\subsection{Radio Continuum}
The area containing W50 was observed for six days with the ATCA \citep{Farnes2017}. The CABB system \citep{Wilson2011} was used for the observations, which provided an instantaneous bandwidth of 2 GHz centered at 2.1 GHz with 2048 spectral channels. Three different types of array configuration were used (H75, H168, and H214). The mosaic provides a very large field of view. The observations were performed on three separate occasions: 22 hr in H75 on 2013 July 05 and 06, 22 hr in H168 on 2013 August 19 and 20, and 22~hr in H214 on 2013 September 15 and 17. PKS1934-638 and J1859+129 were used as the flux calibrator and the phase calibrator respectively. A $3^{\circ}$$\times$$2^{\circ}$ mosaic consisted of 198 pointings arranged in a hexagonal pattern. A summary of the data is provided in Table \ref{tb:1}. 
\par  The polarization leakage and polarization angle calibration were based upon the techniques developed in \citet{Schnitzeler2011} and \citet{Anderson2016}. In order to calibrate these specific observations in MIRIAD, a standard initial calibration was performed to correct for the effects of the bandpass and the complex antenna gains using the tasks {\tt mfcal} $\rightarrow$ {\tt gpcal} $\rightarrow$ {\tt gpcopy} to transfer solutions from the flux calibrator (PKS1934-638) to the phase calibrator (J1859+129). The frequency-dependent instrumental leakages were then calculated in 16 bins across the full bandwidth using {\tt gpcal} and the phase calibrator -- which was observed over a large range in parallactic angle. The absolute flux-scale was bootstrapped using {\tt gpboot}, and the final solutions then linearly interpolated across the 198 pointings of the target W50. All of the sources (including calibrator scans and target pointings) were then gently flagged (being careful to avoid over-flagging) before iteratively performing a new calibration loop (which provided improved solutions due to the removal of bad data). Once flagging was complete, a final calibration loop was then performed to provide the complete dataset. Note that electric-vector polarization angle (EVPA) calibration is not directly performed at the ATCA, but rather the EVPA calibration is provided from an injected XY phase calibration signal. This can calibrate the EVPA to much greater accuracy than is typically achieved with circular feeds and a calibrator source.
\par The Common Astronomy Software Applications (CASA) package was used to create a band-averaged image of Stokes $I$, and MIRIAD was used to create Stokes $I$, $Q$, and $U$ image datacubes. The primary beam effects were considered in both CASA and MIRIAD. Stokes $I$, $Q$, and $U$ datacubes were made every 10 and 100 MHz in frequency using multi-frequency synthesis. Polarized intensity datacubes were also made using task {\tt IMPOL}. 
\par We used the datacubes of Stokes $I$, $Q$, $U$ and polarized intensity every 10 and 100 MHz. In order to make maps of total intensity, polarized intensity, and magnetic field, we used the Astronomical Image Processing System (AIPS). We calculated the polarization angle using task {\tt COMB}. Maps were plotted using task {\tt PCNTR}.
\par W50 is located near the Galactic plane, and the surrounding area is filled with diffuse emission. This makes it challenging to estimate the noise values of each Stokes parameter. We therefore selected a large region to estimate the rms values of Stokes $I$, $Q$, and $U$, and in which W50 and dominant point sources are not included. Nevertheless, the rms values have some uncertainty. On the other hand, to estimate the rms value of polarized intensity, we used the following approach. In general, the thermal noise has the following relationship:
\begin{equation}
\sigma^{{band}}_{QU}=\frac{\sigma^{{sc}}_{QU}}{\sqrt{\textrm{number~of~channels}}},
\end{equation}
where $\sigma^{{band}}_{QU}$ is the band-averaged rms noise, and $\sigma^{{sc}}_{QU}$ is the rms noise in a single freqency channel. We can therefore estimate the single-channel noise in polarized intensity with the following equation:
\begin{equation}
\sigma_{PI} \approx \sigma^{{sc}}_{QU} = \sigma^{{band}}_{QU}\sqrt{\textrm{number~of~channels}}.
\end{equation}
\citet{Farnes2017} measured $\sigma^{{band}}_{QU}$ in the high Faraday depth regions of the Faraday cube to be 0.125~mJy~beam$^{-1}$~rmsf$^{-1}$, and we have adopted this value. \footnote{The unit rmsf is from the ``Rotation Measure Spread Function", which is the equivalent to the point spread function in Faraday space. For further details and the technical background to RM Synthesis, please see \citet{Brentjens2005}.} Using Eq. 2, the rms noise in polarized intensity for 10 and 100 MHz channels can be calculated to be 1.53 and 0.5 mJy/beam respectively (see Table \ref{table2}).

\begin{figure*}[htbp]
 \begin{center}
   \includegraphics[width=16cm]{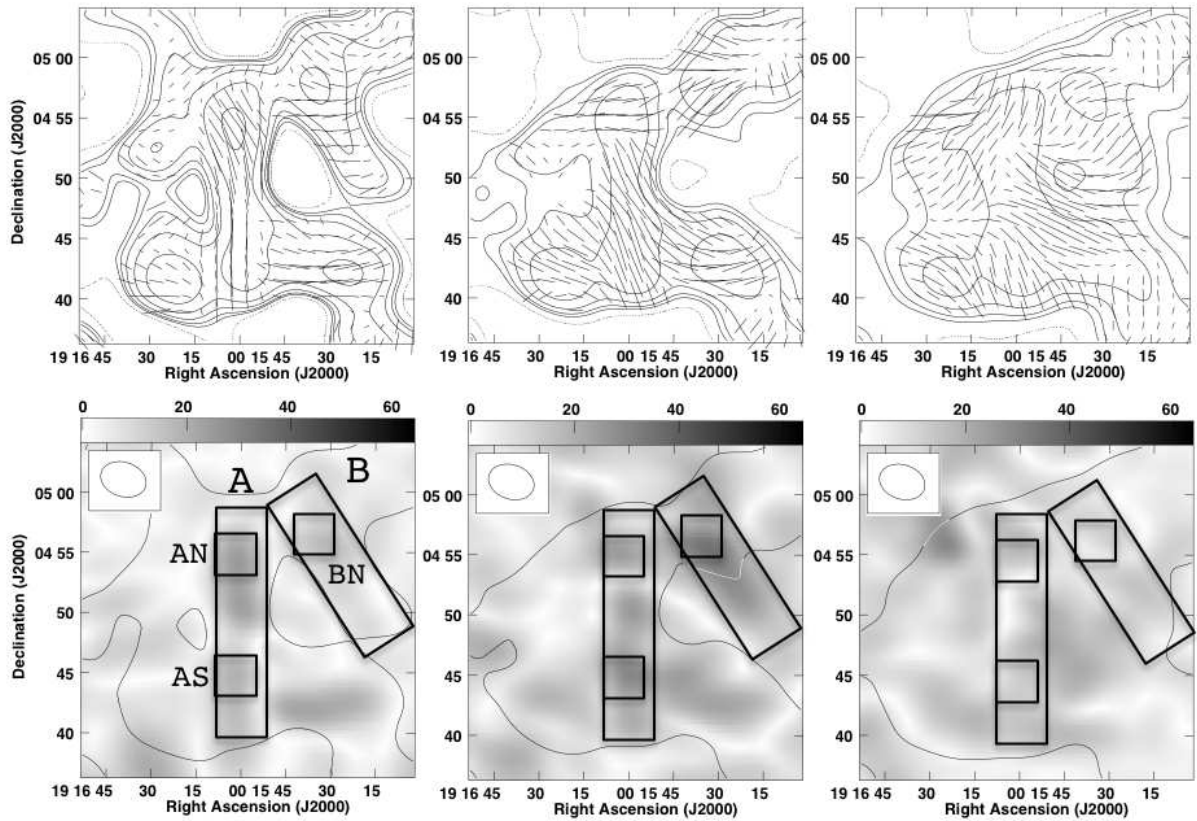}
  \caption{\textbf{(Top left)} An image of W50 at 3.0~GHz from the 100~MHz channel datacube. The white box indicates the region that we analyzed. \textbf{(Top right)} A map indicating specific structures relative to the contours of total radio intensity at 3.0~GHz. \textbf{(Middle row)} Maps of the total intensity (contours) and magnetic field lines (vectors) at three frequencies (3.0~GHz, 2.1~GHz, 1.4~GHz) from the 100~MHz channel datacube. The contours are drawn at the rms noise of total intensity (mJy/beam)$\times$(-1, 1, 2, 4, 8, 16, 32, 64, 128). The magnetic field vectors are only shown at intervals of every five pixels for improved clarity. \textbf{(Bottom row)} The grayscale indicates the polarized intensity. The grayscale range is 0.0 -- 63.3~mJy/beam. The solid line is a contour of the total radio intensity at an rms noise of total intensity (mJy/beam)$\times$1. The top left inset represents the beam size.}\label{figure1}
 \end{center}
\end{figure*}

\begin{figure*}[htbp]
 \begin{center}
   \includegraphics[width=16cm]{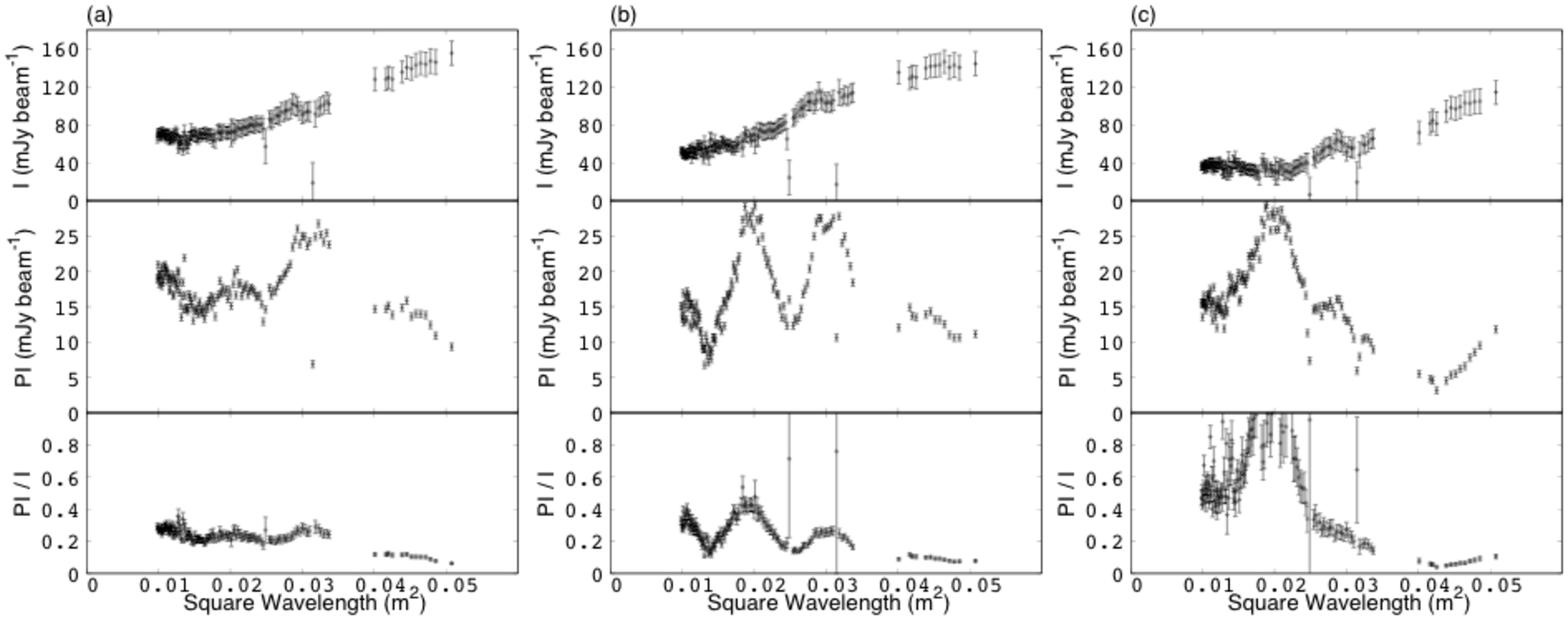}
  \caption{Examples of $I$ (top row), $PI$ (middle row) and $PI/I$ (bottom row) versus $\lambda$$^{2}$ for various locations within the eastern region of W50. (a), (b) and (c) show the results for the regions AN, AS, and BN respectively.}\label{figure2}
 \end{center}
\end{figure*}

\subsection{ \textit{XMM-Newton} Data and Reduction}\label{sec:reduction_XMM} 
\textit{XMM-Newton} was used to observe W50 on 2004 Oct 4 for an exposure of 31.3 ks (OBSID: 0075140501). 
The SAS v13.5 and the built-in extended source analysis software (ESAS) were used to process and calibrate the data obtained with the \textit{XMM-Newton} European Photon Imaging Camera (EPIC).  
Following a standard procedure, the MOS raw data were created by {\tt emchain}, and the light curves were extracted and screened for time-variable background components by the {\tt mos-filter} task.

\section{Results}
As mentioned in Section 1, it is suggested that the SS433 jet reaches the eastern edge of W50 and has made the radio filament and the helical pattern. An X-ray ring has also been detected and overlaps with part of the radio filament. In order to reveal magnetic fields of these specific structures, we performed a radio polarization analysis. We show the distributions of total intensity, polarized intensity, and magnetic vector angle at various frequencies in Section 3.1. Maps of the RM distribution and of the intrinsic magnetic field vectors are shown in Section 3.2.
\subsection{The Distribution of Observational Properties at Radio Frequencies}
The top left panel of Figure \ref{figure1} shows the whole structure of W50 at 3.0~GHz as plotted by using the 100 MHz channel datacube. The white box indicates the eastern-ear region that we focus on in this paper. The top right panel of Figure \ref{figure1} indicates various specific structures: the X-ray ring \citep{Brinkmann2007}, a part of the helical structure \citep{Dubner1998}, the filamentary structure which has previously been identified as the terminal shock, and the bow shock that is formed at the head of the jet. The other six panels of Figure \ref{figure1} show newly derived contours for the total radio intensity (middle row) and the polarized radio intensity (bottom row) in the eastern-ear region. Three different frequencies (3.0, 2.1, and 1.4 GHz respectively) are shown from left to right. The line segments superposed in Figure \ref{figure1} indicate magnetic vectors, where the magnetic vector angle is defined as $\textrm{EVPA} + 90^{\circ}$. The total intensity noise values listed in Table \ref{table2} were used to determine the contour intervals. To simplify our presentation, we define five regions as shown in Figure \ref{figure1} (bottom).
\par The most prominent feature in Figure \ref{figure1} is the bright north--south filamentary structure, at RA: $19^{h}16^{m}00^{s}$ and labelled region A. Some previous studies have suggested that this structure corresponds to the terminal shock \citep{Brinkmann1996,Safi-Harb1997,Safi-Harb1999,Farnes2017}. We can see that the magnetic vectors at 3.0~GHz broadly align with region A and that the vectors rotate as the frequency decreases. The rotation in the northern part of the filament differs from that in the southern part, suggesting different RMs in these two regions. We determine the RMs and the intrinsic magnetic field vectors in the next subsection 3.2.
\par There are also features seen in total intensity at 3.0~GHz. For example, the filament and the structure to the western-side both seem to be associated with the X-ray ring-like structure (See Section 4.3). However, it is difficult to distinguish the structures at lower frequencies. 
\par The polarized intensity in regions A and B are higher than in other areas, and there are two areas in region A which have high polarized intensity values at 3.0~GHz. At 2.1~GHz, the polarized intensities become higher than those at 3.0~GHz in both the southern part of region A and in region B. These structures are blurred by the limited angular resolution and it is difficult to identify them at 1.4~GHz.
\par In order to study the variations of the total and polarized intensities in detail, we divided regions A and B into smaller parts with a size that is roughly consistent with the beam size. We then plotted the $I$, the $PI$ and the $PI$ / $I$ versus $\lambda$$^{2}$ relations in each part, where $I$ is the total intensity, $PI$ is the polarized intensity, $PI$ / $I$ is the fractional polarization, and $\lambda$ is the wavelength.
\par We find that there are roughly three types of spectra in regions A and B. Figure 2(a) shows representative spectra in region AN. The total intensity increases in wavelength, while the polarized intensity is peaked at 18~cm (i.e. 0.032~m$^{2}$) and the fractional polarization gradually decreases in wavelength. Figure 2(b) shows typical spectra in region AS. The spectra are similar to those of the northern part, except that there is another peak at 14~cm (i.e. 0.019~m$^{2}$) in the polarized intensity. Overall, a decrement in the fractional polarization suggests the presence of depolarizing media, although confirmation of this requires further studies. Finally, Fig. 2(c) shows characteristic spectra seen in region BN. The total intensity is flattering at $\lambda <$ 14~cm and increases in wavelength. The polarized intensity has one peak at 14~cm. The fractional polarization shows a complicated variation, and ``repolarization" is occurring with the PI/I values increasing at $\lambda \sim$ 21~cm (i.e. 0.044~m$^{2}$).

\subsection{RM and Intrinsic Magnetic Field Vector Maps}
In order to know the direction and the strength of the magnetic field, we derived maps of the RM and the intrinsic magnetic field vectors. We selected 114 regions which cover the eastern-ear region, and whose size are roughly consistent with the beam size. We plotted the EVPA versus $\lambda^{2}$. We added or subtracted $n\pi$ in order to unwrap any $n\pi$ ambiguity in the polarization angle, and then fitted the EVPAs with a linear function using the least squares method. We derived the best fit which has the smallest chi-squared value, and determined the RM (rad/m$^{2}$) and the intrinsic polarization angle (radians) from the slope and the intercept of the best-fit linear function, respectively. The angle of the intrinsic magnetic field was defined as the intrinsic polarization angle + $\pi$/2. 
\par
\begin{figure*}[t]
 \begin{center}
    \includegraphics[width=16cm]{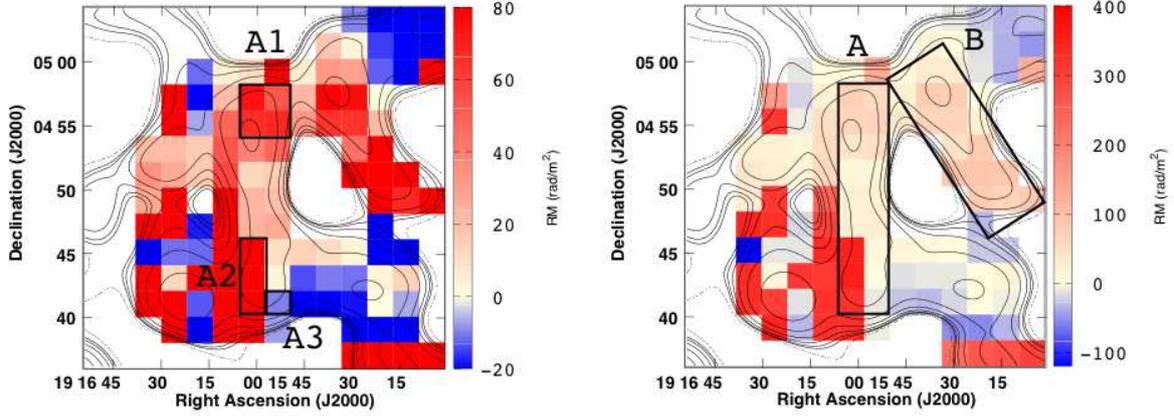}
  \caption{\textbf{(Left)} A RM map, with the pseudocolor scale ranging from -20 to +80~rad/m$^{2}$. Total intensity radio contours are also shown at 3.0~GHz. A1 -- A3 are parts of the terminal shock. \textbf{(Right)} A RM map of the same region as the left panel. However, the pseudocolor scale is ranging from -100 to +400~rad/m$^{2}$.}\label{figure3}
 \end{center}
\end{figure*}

 The derived RMs for the 114 sections in the eastern region are shown in Figure \ref{figure3}. \citet{Farnes2017} estimated the foreground Faraday rotation using the RMs of pulsars to be +15 $\pm$30 rad/m$^{2}$ and regarded it consistent with zero foreground RM, and we adopted the foreground RM $\sim$ 0 rad/m$^{2}$. This map is similar to the RM map shown in \citet{Farnes2017} (See Figures 17 and 18 of their paper). However, there are some deviations due to the different ways the RMs were calculated. For example, the RM at (RA, DEC) = ($19^{h}16^{m}17.77^{s}$, $+04^{\circ}40^{\prime}12.43^{\prime\prime}$) in Figure \ref{figure3} is much lower than the value from \citet{Farnes2017}. We further note that the RM values of the pixels are also sensitive to the actual gridding, e.g.\ the number of pixels and the grid coordinates.
\par
\begin{figure}[t]
 \begin{center}
    \includegraphics[width=8cm]{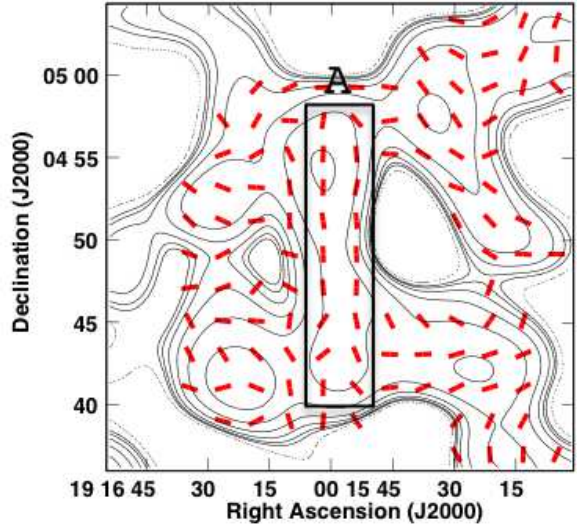}
  \caption{A map of the intrinsic magnetic field vectors (red lines) with contours of the total intensity at 3.0~GHz (black contours). Note that the length of the vectors does not correspond to magnetic field strength.}\label{figure4}
  \end{center}
\end{figure}

 Figure \ref{figure4} shows a map of the intrinsic magnetic field vectors overlaid with contours from the total intensity image. We find that the magnetic vectors are oriented in the north--south direction in the filament corresponding to region A, which is thought to be the terminal shock in previous works \citep{Brinkmann1996,Safi-Harb1997,Safi-Harb1999,Farnes2017}. We also see an approximate alignment between the total intensity structure and the intrinsic magnetic field vectors in the surface of the eastern region of W50, which is considered to be the bow shock (See Figure \ref{figure1}, top right panel). Note that the vectors on the precession axis of the SS433 jet are perpendicular to the bow shock. We discuss the magnetic field structure of the bow shock in Section 4.2.

\section{Discussion}
In this Section, we discuss specific individual structures in the eastern ear of W50. The first point that we consider is the structure at the location of the terminal shock. In order to clarify this, we apply Faraday tomography to the terminal shock region. The next point that we consider is the correlation between the X-ray ring and the radio filament. Finally, we propose a model of the eastern edge of W50 based on our results.

\subsection{Different Characteristics at the Terminal Shock : region A}
As introduced earlier in this paper, some previous works suggest that the north--south filament in the eastern ear corresponds to the terminal shock of the SS433 jet. However, in Section 3.1 we found that the northern and southern parts of the filament have different trends in total intensity, polarized intensity, and magnetic vectors relative to each other. Some previous observations have actually pointed out substructures inside the filament. For example, \citet{Dubner1998} claimed that the total radio intensity in the southern part is much less bright than in the northern part, and this trend can also be seen in our total intensity contours (Figure \ref{figure1}, top panel). Furthermore, \citet{Brinkmann2007} showed that the southern part is not bright in the X-ray. They discovered a ring-like structure (See Section 4.3) that includes the regions associated with both the northern part of region A as well as region B. They proposed that the ring corresponds to the terminal shock of the SS433 jet. These and our observations suggest that region A can be divided into at least two separate regions: with distinct northern and southern parts.

\par In order to further study region A, we checked the distribution of RM. Significantly, the RM values are 323 -- 355 rad/m$^{2}$ in region A2, which is the southeast part of region A, and much larger than the values in region A1 which is the north part of region A (53 -- 74 rad/m$^{2}$, see Figure \ref{figure3}). Also, the RM values decrease from region A1 to region A3 which lies the southwest part of region A: the line-of-sight components of the magnetic fields in region A1 and region A3 point in opposite directions relative to each other. Therefore, our RM map indicates that there are significantly different magnetic field structures between the northern and southern parts of region A.

\begin{figure*}[t]
 \begin{center}
    \includegraphics[width=16cm]{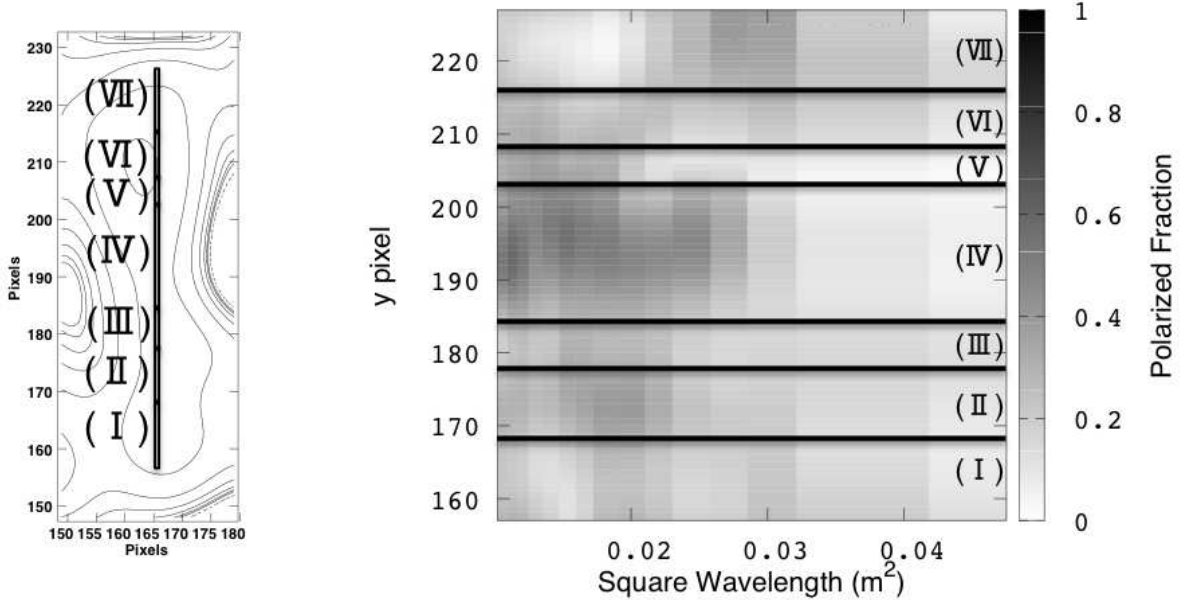}
  \caption{\textbf{(Left)} A plot showing the sectors in which we carried out Faraday Tomography. The contours of total intensity at 3.0~GHz are also shown. These regions are all located along the filamentary structure in region A. \textbf{(Right)} The fractional polarization as a function of the square of the wavelength. The grayscale indicates the fractional polarization. The solid lines represent the boundary of each sector.}\label{figure5}
   \end{center}
\end{figure*}

\begin{figure*}[htbp]
 \begin{center}
    \includegraphics[width=14cm]{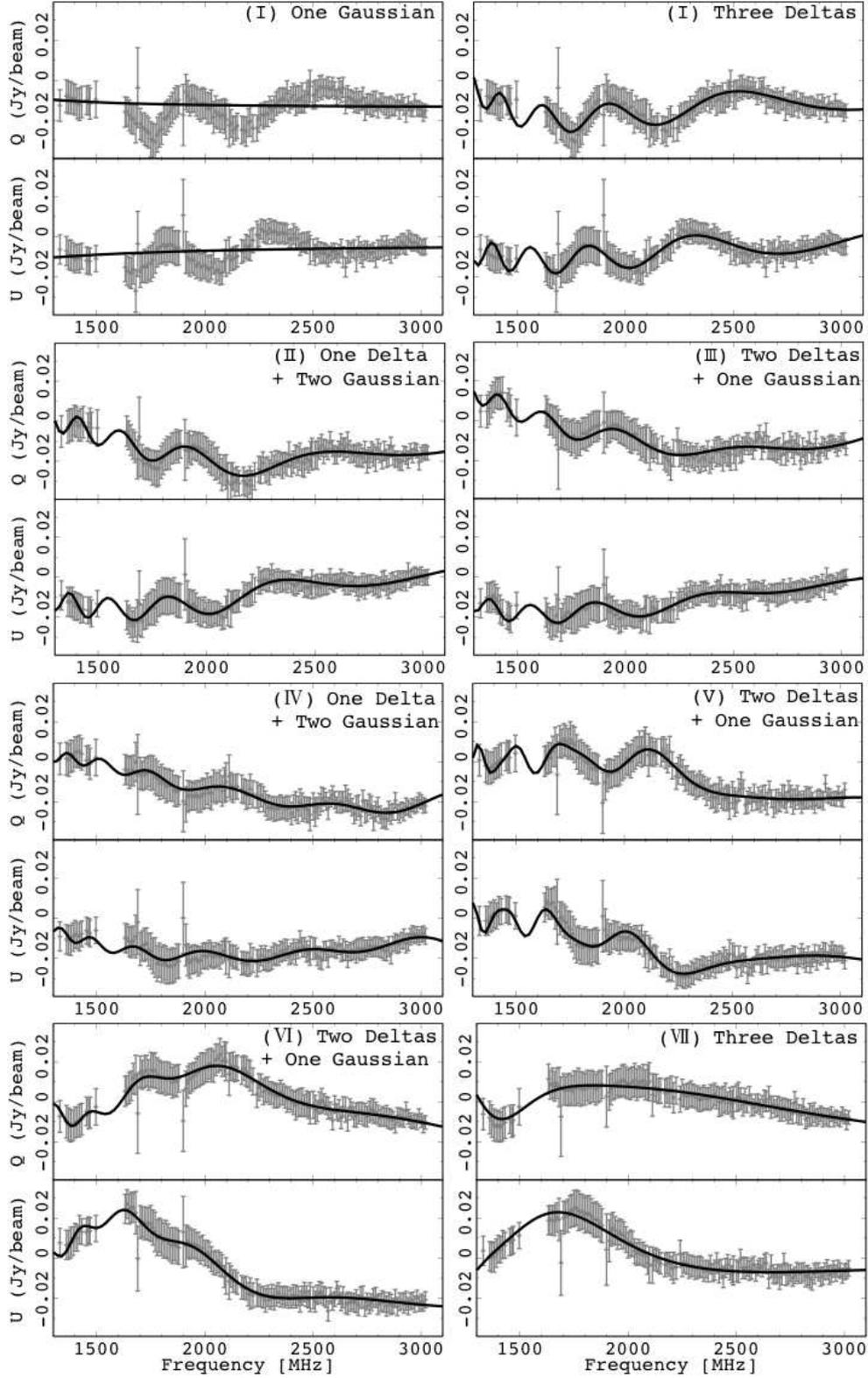}
  \caption{Plots of the Stokes $Q$ and $U$ spectra and the associated fits to the data. The black lines indicate the best fitting model, while the data and associated uncertainties are shown in grey. All fitted models indicate the best fitting model, with the exception of the plot in the top left corner labelled ``(I) One Gaussian''.}\label{figure6}
  \end{center}
\end{figure*}

\begin{longtable}{llllllll}
 \caption{The results of the $QU$-fitting and the values of the best fitting model and 1$\sigma$ confidence regions for each model parameter.}\label{table3}
\hline
Region &y pixel&Model&BIC&$\phi$ (rad/m$^{2}$)&Amp. (mJy)&$\chi_{0}$ (rad)&$\sigma$ (rad/m$^{2}$)\\
\hline
\endhead
\hline
\endfoot
\hline
\multicolumn{8}{}{\hbox to 0pt{\parbox{180mm}}}
\endlastfoot
I & 157--167 & Three Deltas  & 102 & 314$^{+8.6} _{-6.3}$ & 6.50$^{+0.49} _{-0.57}$ & -1.4$^{+0.2} _{-0.2}$ & \\
 & & & & -37.4$^{+13} _{-6.5}$ & 8.07$_{-1.6} ^{+1.6}$ & -0.13$_{-0.4} ^{+0.2}$ & \\
 & & & & 24.2$_{-5.2} ^{+6.5}$ & 14.5$_{-2.0} ^{+1.1}$ & 1.1$_{-0.2} ^{+0.2}$ &  \\
I\hspace{-.1em}I & 168--177 & One Delta  & 99.9 & 302$_{-9.4} ^{+8.8}$ & 5.48$_{-0.64} ^{+0.82}$ & -1.1$_{-0.2} ^{+0.2}$ & \\
 & & + Two Gaussian & & 49.2 $_{-47} ^{+46}$ & 43.6$_{-14} ^{+21}$ & -0.21$_{-0.4} ^{+0.5}$ & 67.9$_{-7.2} ^{+20}$ \\
 & & & & 16.2$_{-2.0} ^{+4.3}$ & 28.2$_{-3.8} ^{+1.2}$ & 1.5$_{-0.1} ^{+0.1}$ & 12.7$_{-2.8} ^{+1.3}$ \\
I\hspace{-.1em}I\hspace{-.1em}I & 178--183 & Two Deltas & 97.3 & 293$_{-12} ^{+16}$ & 4.58$_{-0.98} ^{+0.068}$ & -0.61$_{-0.4} ^{+0.2}$ & \\
 & & + One Gaussian & & 22.5$_{-19} ^{+25}$ & 18.6$_{-1.2} ^{+0.91}$ & -1.6$_{+3} ^{+3}$ & \\
 & & & & 229$_{-110} ^{+46}$ & 82.6$_{-50} ^{-3.4}$ & 1.4$_{-0.4} ^{+0.2}$ & 98.9$_{-17} ^{+2.1}$ \\
I\hspace{-.1em}V & 184--202 & One Delta & 105 & 337 $_{-64} ^{+67}$ & 2.63 $_{-1.2} ^{+0.52}$ & -0.59$_{-0.6} ^{+0.7}$ & \\
 & & +Two Gaussian & & 16.0$_{-2.5} ^{+3.0}$ & 30.5$_{-1.9} ^{+1.8}$ & -1.5$_{-0.1} ^{+0.06}$ & 16.4$_{-1.3} ^{+1.9}$ \\
 & & & & 564$_{-146} ^{+185}$ & 44.4$_{-32} ^{+5.6}$ & 1.3$_{-1} ^{+0.2}$ & 93.9$_{-12} ^{+6.9}$ \\
V & 203--207 & Two Deltas  & 104 & 504 $_{-310} ^{-94}$ & 1.88$_{+3.4} ^{+4.9}$ & -0.69$_{+0.8} ^{+2}$ & \\
 & & + One Gaussian & & 299$_{-77} ^{+34}$ & 6.43$_{-5.6} ^{-4.1}$ & 0.40$_{-0.3} ^{+0.9}$ & \\
 & & & & 32.0$_{-4.1} ^{+3.9}$ & 39.6$_{-1.6} ^{+2.6}$ & -1.5$_{-0.09} ^{+0.08}$ & 31.8$_{-1.3} ^{+2.1}$ \\
V\hspace{-.1em}I & 208--215 & Two Deltas & 95.5 & 296$_{-15} ^{+19}$ & 3.27$_{-0.58} ^{+0.91}$ & 0.21$_{-0.3} ^{+0.2}$ & \\
 & & + One Gaussian & & -21.3$_{-14} ^{+13}$ & 6.21$_{-1.3} ^{+1.4}$ & -1.4$_{-0.1} ^{+0.4}$ & \\
 & & & & 64.9$_{-5.4} ^{+3.5}$ & 26.6$_{-3.4} ^{+3.8}$ & 1.6$_{-0.2} ^{+0.01}$ & 14.9$_{-3.4} ^{+4.0}$ \\
V\hspace{-.1em}I\hspace{-.1em}I & 216--227 & Three Deltas & 93.3 & 69.8 $_{-190} ^{+11}$ & 12.9$_{-6.2} ^{-3.3}$ & 1.6$_{-0.4} ^{+0.07}$ & \\
 & & & & -140$_{-10} ^{+200}$ & 2.28$_{+4.1} ^{+7.3}$ & -1.4$_{+2} ^{+3}$ & \\
 & & & & -25.4$_{+0.72} ^{+160}$ & 9.04$_{+1.8} ^{+4.6}$ & 1.5$_{-0.3} ^{+0.08}$ & \\
\end{longtable}

\par In order to extract further information on the RM, we performed Faraday Tomography. We selected a $1\times71$ pixel box along the filament in region A (RA: $19^{h}15^{m}58.704^{s}$ -- $19^{h}15^{m}58.794^{s}$, DEC: $04^{\circ}41^{\prime}12.83^{\prime\prime}$ -- $04^{\circ}58^{\prime}42.83^{\prime\prime}$) and divided the structure into seven sectors according to the apparent characteristics of the polarized fraction, which we define as
\begin{equation}
p = \frac{{\sqrt{Q^{2}+U^{2}}}}{I},
\end{equation}
with Stokes $I$, $Q$, and $U$ (Figure \ref{figure5}).
\par We employed the Markov chain Monte Carlo $QU$-fit method (\cite{Sullivan2012}; \cite{Ideguchi2014}; \cite{Ozawa2015} and references therein). The fitting formulas that we assumed are a delta function and a Gaussian. Both functions consisted of a Faraday depth $\phi$, an amplitude, and an intrinsic polarization angle $\chi$$_{0}$. The dispersion $\sigma$ was also included as an additional parameter for the Gaussian function. Nine models were then applied in each region: One Delta, One Gaussian, Two Deltas, One Delta + One Gaussian, Two Gaussians, Three Deltas, Two Deltas + One Gaussian, One Delta + Two Gaussians, and Three Gaussians.
\par Estimating the errors in Stokes $Q$ and $U$ is challenging because of diffuse emission from the Galaxy. We applied the rms values of Stokes $Q$ and $U$ during the $QU$-fitting. As a result, the reduced chi-squared ($\chi_{r}^2$) is less than unity in most model fits. We plotted the Stokes $Q$ and $U$ spectra and the associated fits in which $\chi_{r}^2$ are near to unity, and we found that the $\chi_{r}^2$ are not a suitable indicator to confirm the reproducibility of the fitting 
(Figure \ref{figure6} -- I -- One Gaussian, where $\chi_{r}^2$ is near unity, but the model fit to the spectrum is clearly poor). We therefore compared the spectra of Stokes $Q$ and $U$ and the results of $QU$-fitting. If there were multiple well-fitting models, we chose the one which has the lowest Bayesian information criterion (BIC) value.
\par The seven panels in Figure \ref{figure6} show the plots of the Stokes $Q$ and $U$ spectra and the associated best fits. In all of the sectors, the three-component models provide reasonable fits. This result suggests that there are complex structures along the line of sight and/or within the beam area. The best fit parameters are listed in Table~\ref{table3}. All sectors have one or two, large, several hundred rad/m$^2$, Faraday depth components. This is consistent with the foreground estimates in this area in \citet{Farnes2017}. They estimated the typical foreground of the Galactic emission surrounding the ear to be $\sim$ 230 rad/m$^{2}$, with negative RMs up to -100 rad/m$^{2}$ in the north and positive RMs up to +670 rad/m$^{2}$ in the south. Except these large Faraday depth components, only one component is associated with W50 in sectors (I\hspace{-.1em}I\hspace{-.1em}I), (I\hspace{-.1em}V), and (V). On the other hand, there are two components that are related to W50 in sectors (I), (I\hspace{-.1em}I), (V\hspace{-.1em}I), and (V\hspace{-.1em}I\hspace{-.1em}I) respectively. We therefore suggest that the filament can be divided into at least two separate northern and southern parts around sectors (I\hspace{-.1em}I\hspace{-.1em}I) -- (V). Note that a possibility of either or both components, which are detected in sectors (I), (I\hspace{-.1em}I), (V\hspace{-.1em}I), and (V\hspace{-.1em}I\hspace{-.1em}I), resulting from different structures, cannot be dropped. Higher angular resolution and wider frequency coverage observations are needed to identify more detailed structures toward these sectors.

\subsection{Magnetic Field Structure of the Bow Shock}
As mentioned in Section 3.2, the magnetic field of the bow shock is perpendicular to the shock front on the precession axis of the SS433 jet. We consider two possibilities for how such a magnetic field structure may be created. First, it might be caused by the limited low angular resolution of the observations. \citet{Reynoso2013} shows the intrinsic magnetic field vector map of supernova remnant, SN 1006, as observed with the VLA and ATCA (See Figure 3 of \cite{Reynoso2013}). In their figure, the intrinsic magnetic fields are parallel to the shock. On the other hand, the magnetic fields perpendicular to the shock mostly dominate down-stream of the flow from the shock front. Higher resolution observations of the eastern ear of W50 would be able to research whether the magnetic field vectors are parallel to the bow shock on the precession axis.
\par Second, we consider the jet model which could explain such a magnetic field structure. \citet{Kudoh2002} conducted MHD numerical simulations of a jet in which a weak localized poloidal magnetic field is initially embedded. In this simulation, the magnetic fields are parallel to the bow-shock except near the jet axis (See Figure 2 of \cite{Kudoh2002}). It is similar to the magnetic field structure of Figure \ref{figure4}. Of course it is an idealized model, and the real source will have more complicated structure.

\begin{figure}[htbp]
 \begin{center}
    \includegraphics[width=8cm]{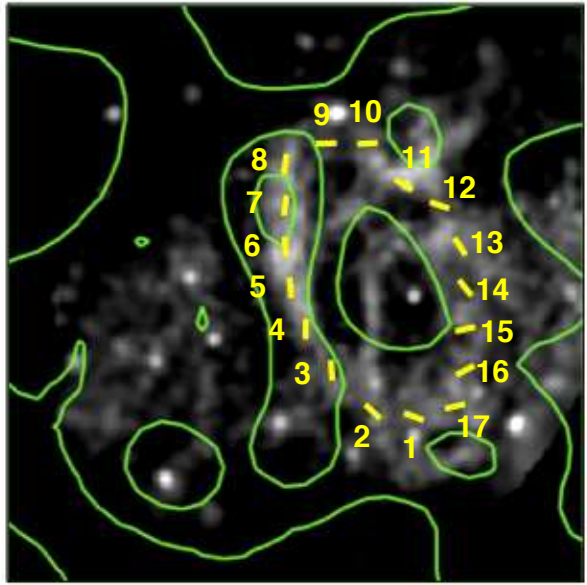}
  \caption{A plot showing the X-ray image from \textit{XMM-Newton} in grayscale. Energy band is 0.4 -- 1.3 keV. The yellow lines are the intrinsic magnetic field vectors. Total intensity radio contours at 3.0~GHz are shown in green. The numbers corresponds to the IDs of the vectors which are adopted to the horizontal axis of figure \ref{figure8}.}\label{figure7}
  \end{center}
\end{figure}

\subsection{Comparison with the X-ray Ring}
If it is the case that the X-ray ring-like structure corresponds to the terminal shock of the SS433 jet \citep{Brinkmann2007}, one can expect that the northern part of region A is continuously connected to region B. Figure~\ref{figure7} displays the intrinsic magnetic field vectors overlaid on the X-ray ring-like structure. It is clear that the majority of vectors are oriented along the ring. On the other hand, the total intensity map indicates that there is a gap (a region with weak radio emission) between regions A and B (see also \cite{Dubner1998}). 

\par
\begin{figure*}[t]
 \begin{center}
    \includegraphics[width=16cm]{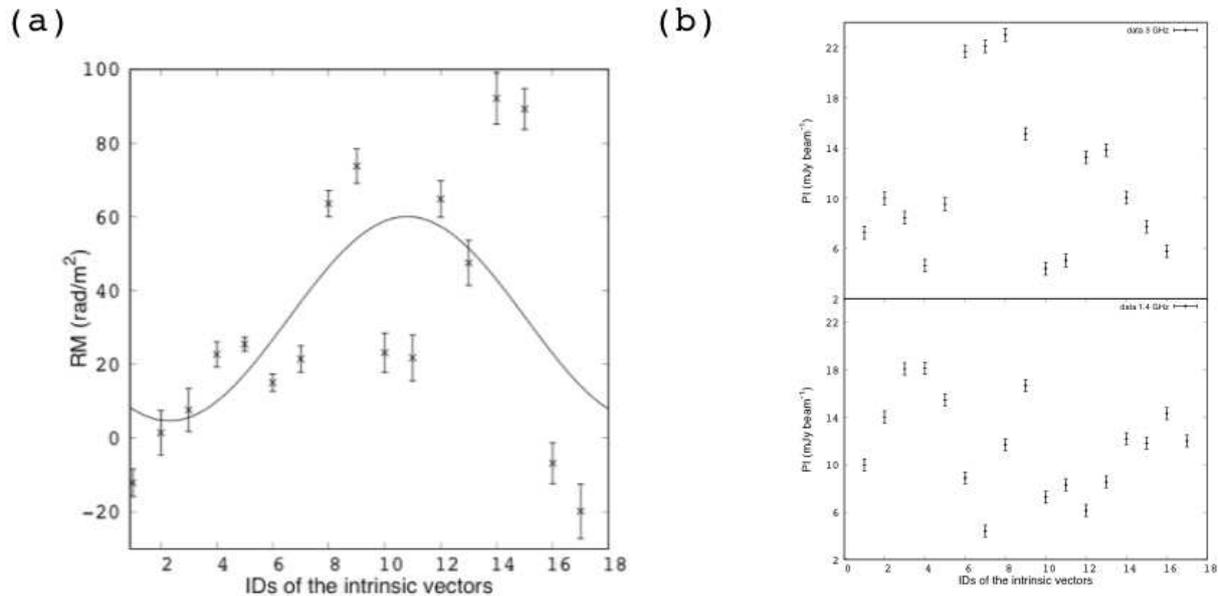}
  \caption{(a) The variation of RMs along the ring. The solid line represents the result of the sine function fitting. Position numbers 4 -- 9 and 11 -- 16 correspond to the northern part structure of region A and the structure of region B. (b) The variation of the polarized intensity along the ring. The top panel shows the result at 3.0~GHz, and the bottom panel shows the result at 1.4~GHz.}\label{figure8}
   \end{center}
\end{figure*}

 To explore this hypothesis, we analyzed the region further. For the purpose of simplicity, we assume: (1) that the radius of the ring is equal to the radius of the eastern ear, (2) that regions A and B are located in the far side and the near side respectively in line with \citet{Brinkmann2007}, and (3) that the observed magnetic fields consist of a radial component and a circumferential component. With these assumptions, if the ring is continuous, the variation of RMs should follow a sine curve along the ring. Figure~\ref{figure8}(a) shows our result, where the numbers on Figure~\ref{figure7} show the IDs of the intrinsic vectors which are adopted on the horizontal axis of Figure~\ref{figure8}. The Figure clearly indicates that the RMs do not follow the best-fit sine curve, particularly between sectors 9 and 10, and also 15 and 16. One could argue that some error bars are underestimated. However, this seems unlikely, as both the RMs and the error bars are estimated by considering the relative weight of each data point. This suggests that the error bars are accurate.

\par We next compared the polarized intensity for each of the IDs (Figure~\ref{figure8}b). While the polarized intensity seem very low in some regions, these are larger than the lower limit of $\sim0.5$~mJy/beam. Figure \ref{figure8}(b) suggests that the polarized intensity differ in each region along the ring. We also checked the variation of the polarized intensity as a function of $\lambda^{2}$. At IDs = 7 to 9, the polarized intensity become lower at $\lambda$ = 20 -- 22 cm similar to Figure~\ref{figure2}(a). On the other hand, IDs = 10 to 15 show that the polarized intensity becomes higher similar to Figure~\ref{figure2}(c). Considering these results, it seems that the regions A and B are discontinuous. However, the possibility remains that foreground matter causes extra Faraday rotation that hides the magnetic structure of the ``real" ring. This hypothesis explains the differences in the variation of the RMs and the polarized intensity along the ring. Foreground matter may induce ``repolarization" with the fractional polarization increasing at longer wavelengths, and with the polarized intensity increasing at longer wavelengths in the western part of the ring. In order to explore these possibilities in more detail, high angular resolution observations are needed.

\begin{figure*}[htbp]
 \begin{center}
    \includegraphics[width=16cm]{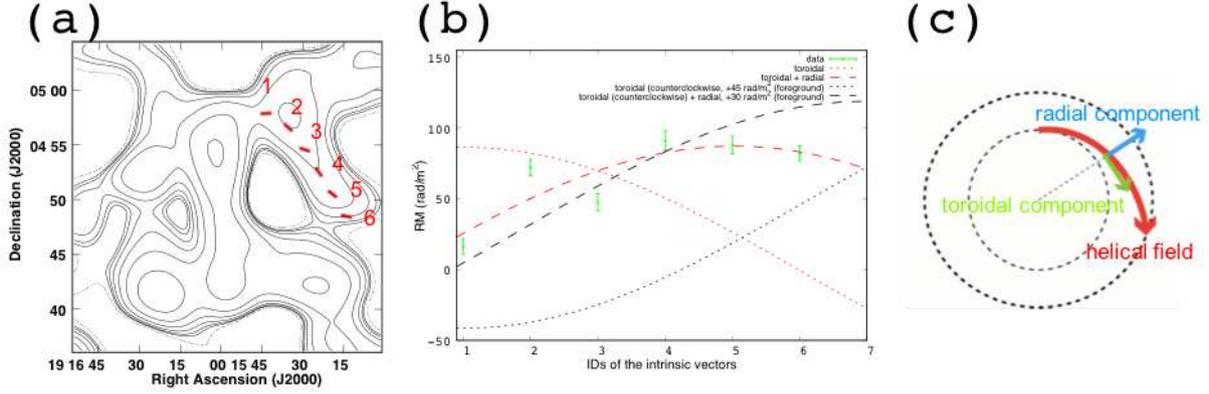}
  \caption{(a) The intrinsic magnetic field vectors (red lines) with total intensity radio contours at 3.0~GHz (black contours). The numbers show the IDs of the vectors which are adopted to the horizontal axis of (b). (b) The variation of the RMs along the structure of region B. The dotted lines represent various toroidal magnetic field models, with the clockwise and counter-clockwise directions being indicated by red and black respectively. Models which only include a toroidal component are indicated by the dotted lines, whereas models that include both a toroidal and a radial magnetic field component are indicated by the dashed lines. The counter-clockwise models also include foreground RMs as indicated in the Figure, whereas the clockwise models include zero foreground RM. (c) A cartoon illustration of the magnetic field model, when assuming a clockwise toroidal component and a radial component.}\label{figure9}
   \end{center}
\end{figure*}

\subsection{Evidence of a Helical Structure Generated by the SS433 Jet : region B}
Although we do not rule out the possibility that regions A and B form a continuous structure, our analysis in Section 4.3 suggests that they are more likely separate, and we will therefore assume that they are independent. \citet{Dubner1998} reported that the eastern ear exhibits a helical pattern. Therefore, we assumed that the structure of region B was a part of the helical pattern and analyzed this structure following this hypothesis. 
\par Figure \ref{figure9}(a) shows the intrinsic magnetic field vectors along region B. The numbers on Figure \ref{figure9}(a) are the position IDs that correspond to the number on the horizontal axis of Figure \ref{figure9}(b). Figure \ref{figure9}(a) indicates that the intrinsic magnetic field vectors extend along the structure. The RMs become larger at higher position IDs from 1 to 4 (Figure 9b). We therefore present a magnetic field model that explains the observed trend of the RMs. We assume that the helical structure coils around the ear with a clockwise rotation as seen from the eastern edge to the center of W50. First, we considered a toroidal-magnetic field.  In this model, the component of the magnetic field along the line of sight gets smaller as the position gets nearer to the axis of the helical pattern, and the RMs therefore also decrease (See the red-dotted line in Figure \ref{figure9}b). Next, we added a magnetic field oriented in the radial direction towards the outside of the eastern ear. The ear width gradually increases from the terminal shock to the helical structure. This expanding structure corresponds to the introducing radial component. The additional radial component therefore does not contradict the structure of the helical pattern. This model provides a good explanation for the RMs of the structure (See the red-dashed line in Figure \ref{figure9}b, and Figure \ref{figure9}c). In this model, the strength ratio (toroidal-magnetic component)~:~(radial-magnetic component) is 1~:~3.
\par We also assumed that the helical structure coils around the ear with counter-clockwise rotation. First, we considered a toroidal magnetic field, without any radial component. In this case, the RMs increase at higher position IDs from 1 to 6. However, all of these values should be negative since the magnetic field vectors are oriented away from us along the line of sight. Even if we assume a foreground RM~=~+45~rad/m$^{2}$, which is the maximum value estimated in previous studies, this model still cannot explain the observations (See the black-dotted line in Figure \ref{figure9}b). Next, we added a radial magnetic component. When the strong radial magnetic fields whose strength is 88 times stronger than the toroidal component are assumed, the resultant distribution would be able to explain the variation of the data. This assumption indicates, however, the magnetic fields are distributed radially and do not distribute helically. These situation might be formed around the terminal shock. Since the helical structure we called is far from the terminal shock, we considered this model should be excluded. Next, we set the strength ratio (toroidal-magnetic component)~:~(radial-magnetic component) to be 1~:~3, and assumed the best fitting foreground RM~=~+30~rad/m$^{2}$. The model is shown in Figure \ref{figure9}(b), as indicated by the black-dashed line. This model seems to provide a reasonable fit to the data. However, the reduced chi-square of this model is calculated to be 20.9. This is worse than the value of 8.24 obtained for the combined clockwise toroidal+radial model with zero foreground RM.

\begin{figure*}[t]
 \begin{center}
    \includegraphics[width=16cm]{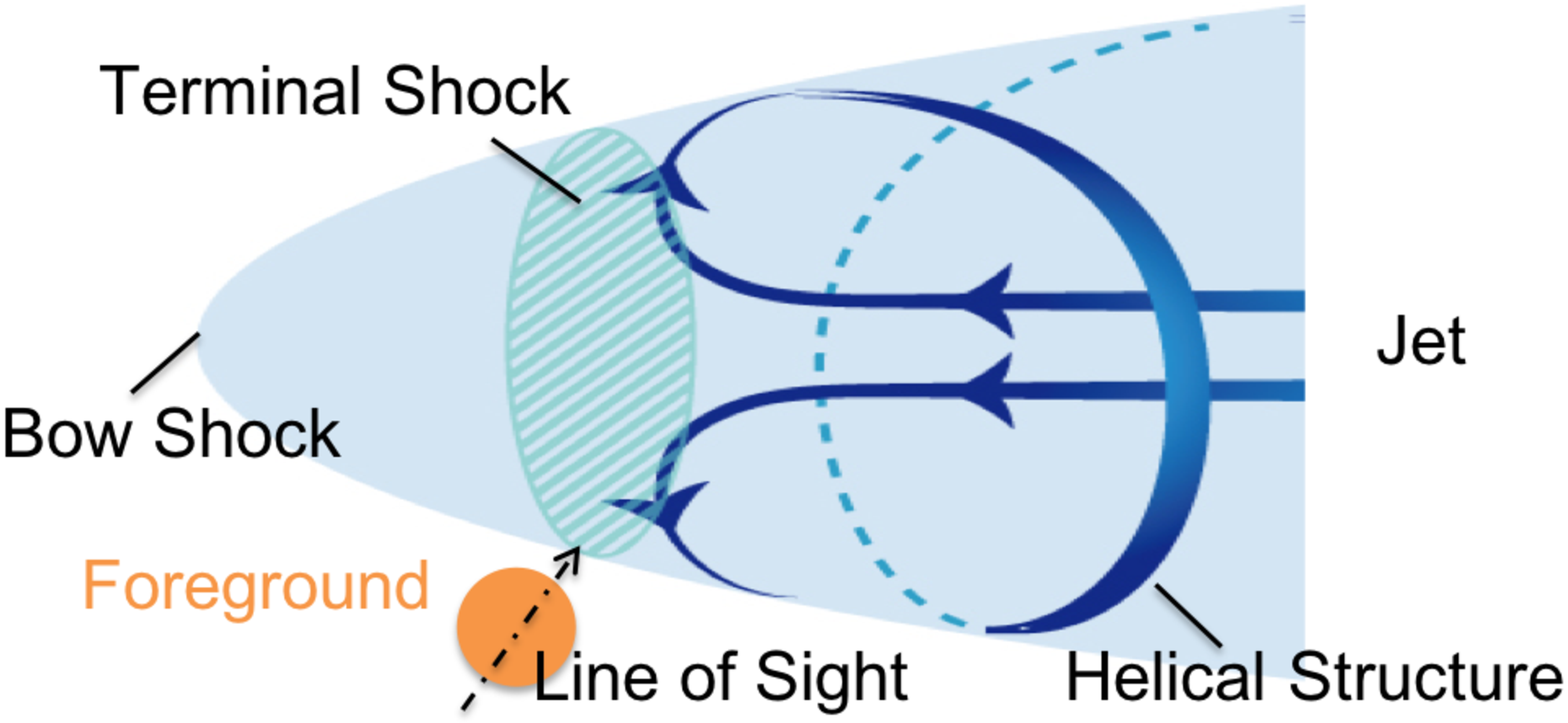}
   \caption{The model of the eastern ear of W50.}\label{figure10}
  \end{center}
 \end{figure*}

\par It is possible to consider various additional models by changing the strength ratio or the foreground RM. Further observations are needed to constrain the detail model. For example, high resolution X-ray observation would be able to provide the thermal electron number density, thereby allowing for inference of the non-thermal electron number density. This would allow for an estimate of the intensity of the magnetic field.
\par Figure 10 shows our proposed model of the helical pattern when we assume the red-dashed line model of Figure \ref{figure9}(b) (clockwise toroidal + radial magnetic field). It also gives a possible process through which the helical pattern could be made. In this model, the SS433 jet goes through W50 and forms both the terminal shock and the bow shock. Matter is pushed away at the terminal shock and reaches the surface of the eastern ear of W50. It still has angular momentum, and coils the ear.
\par \citet{Gabuzda2015} suggested that active galactic nuclei (AGN) jets have gradients of RMs in the direction that is perpendicular to the jet axes. Region B also shows a similar RM gradient, which is almost perpendicular to the jet axis. This may suggest a relation to the structures seen in AGN jets.

\section{Conclusion}
We carried out a polarization analysis of the eastern region of W50 as observed by the ATCA at 1.1 -- 3.1 GHz, and we used the radio data at 1.4 -- 3.0 GHz with 16 and 150 individual frequencies. We analysed the maps of total intensity, polarized intensity, and magnetic field vectors. We also made new maps of the RMs and the intrinsic magnetic field vectors.
\begin{itemize}
	\item The distribution of RMs seems to resemble the result of \citet{Farnes2017}. However, in some areas the values of RM are different to the previous one.
	\item The magnetic vectors are oriented along the bright total intensity filament. This is consistent with the filament cospatially exist with the terminal shock of the jet.
\item The bright total intensity filament in the eastern ear can be divided into at least two parts to the north and south based upon the differences of the RMs and other observations. The results of Faraday Tomography support this hypothesis. The results of Faraday Tomography also suggest that there are complex structures that consist of about three components along the line of sight and/or within the beam area. Observations with higher angular resolution and wider frequency coverage are needed to understand these structures in more detail.
	\item The magnetic field vectors of the bow shock are almost parallel to the shock front, while the vectors are perpendicular to the shock on the precession axis of the SS433 jet. We propose two reasons to explain this. One possibility is the limitation of the spatial resolution. In addition, based on the comparison with the numerical simulation, we propose the other possibility that the observed structure is a real structure.
	\item We analyzed the X-ray ring observed with \textit{XMM-Newton}. It seems that this structure is discontinuous as seen in the radio. However, there is a possibility that a ``real'' continuous ring cannot be fully seen due to foreground intervening magnetised matter.
	\item We analyzed the structure in region B in Figure \ref{figure1}. We considered some models of the magnetic field, and the toroidal+radial magnetic field models seems to provide the best fit to the observational data. Additional observation is needed to limit these models. This structure has an RM gradient. It may therefore have a relation to the RM gradients previously seen in AGN jets.
\end{itemize}

\section*{Acknowledgements}
We are grateful to Drs. K. Tachihara, H. Yamamoto, and H. Sano for useful discussion. We also thank N. Egashira for making the great schematic (Figure \ref{figure10}). We thank the anonymous referee for useful comments and constructive suggestions. The Australia Telescope Compact Array is funded by the Commonwealth of Australia for operation as a National Facility managed by CSIRO. This paper is based on observations obtained with \textit{XMM-Newton}, an ESA science mission with instruments and contributions directly funded by ESA Member States and NASA. This work is financially supported in part by a Grant-in-Aid for Scientific Research (KAKENHI) from JSPS (MM:16H03954, TA:15K17614, 17H01110). H.A.~acknowledges the support of NWO via a Veni grant. SRON is supported financially by NWO, the Netherlands Organization for Scientific Research. A part of this research is supported by Qdai-jump Research Program(PI:MM, 29212).

\end{document}